\documentclass[a4paper,twoside]{jpconf_mod}
\usepackage{graphicx}
\usepackage[compress]{cite}
\usepackage{hyperref,url}
\usepackage{amsmath,amssymb}
\DeclareGraphicsExtensions{.pdf}

\pdfoutput=1 

\def\t#1{\tilde{ #1}}
\def\tev{\textrm{TeV}}
\def\gev{\textrm{GeV}}

\def\ifb{\ensuremath{\textrm{fb}^{-1}}}

\def\W{\ensuremath{W}}

\def\to{\ensuremath{\rightarrow}}

\def\X{\ensuremath{\tilde\chi_1^0}}
\def\x{\ensuremath{\chi}}
\def\mx{\ensuremath{m_{\chi}}}
\def\R{\emph{R}}

\def\pt{\ensuremath{p_{\rm T}}}
\def\et{\ensuremath{E_{\rm T}}}
\def\met{\ensuremath{E_{\rm T}^{\rm miss}}}
\def\mef{\ensuremath{M_{\rm eff}}}

\def\mt{\ensuremath{M_{\rm T}}}

\begin{document}

\title{Overview of searches for dark matter at the LHC}

\author{Vasiliki A Mitsou}

\address{Instituto de F\'isica Corpuscular (IFIC), CSIC -- Universitat de Val\`encia, \\ 
Parc Cient\'ific de la U.V., C/ Catedr\'atico Jos\'e Beltr\'an 2, \\
E-46980 Paterna (Valencia), Spain}

\ead{vasiliki.mitsou@ific.uv.es}

\begin{abstract}
Dark matter remains one of the most puzzling mysteries in Fundamental Physics of our times. Experiments at high-energy physics colliders are expected to shed light to its nature and determine its properties. This review talk focuses on recent searches for dark-matter signatures at the Large Hadron Collider, either within specific theoretical scenarios, such as supersymmetry, or in a model-independent scheme looking for mono-$X$ events arising in WIMP-pair production. 
\end{abstract}

\markright{V~A~Mitsou}

\section{Introduction}\label{sc:intro}

Both Astroparticle and Particle Physics pursue the exploration of the nature of dark matter (DM)~\cite{dm-review}. Among the long list of well-motivated candidates, the most popular particles are \emph{cold} and weakly interacting, typically predicting missing-energy signals at particle colliders. Supersymmetry~\cite{susy-dm} and models with extra dimensions~\cite{ued-dm} are theoretical ideas that inherently provide such a dark matter candidate. High-energy colliders, such as the Large Hadron Collider~\cite{lhc} at CERN, are ideal machines for producing and eventually detecting DM~\cite{ijmpa}. 

The structure of this paper is as follows. Section~\ref{sc:intro} provides a brief introduction to the relevance of colliders, and in particular the LHC experiments ATLAS and CMS, for the production of dark matter. In Section~\ref{sc:monox}, the strategy, methods, and results of the LHC experiments as far as model-independent DM-production is concerned are discussed. In Section~\ref{sc:susy}, the latest results in searches for supersymmetry at the LHC are presented. The paper concludes with a summary and an outlook in Section~\ref{sc:summary}.

\subsection{Dark matter and colliders}\label{sc:dm}

The nature of the dark sector of the Universe constitutes one of the major mysteries in fundamental physics. According to recent observations from anisotropies of the cosmic microwave background made by the Planck mission team~\cite{planck2}, most of our Universe energy budget consists of unknown entities: $\sim\!26.8\%$ is dark matter and $\sim\!68.3\%$ is dark energy, a form of ground-state energy. Dark energy is believed to be responsible for the current-era acceleration of the Universe. Dark matter, on the other hand, is matter inferred to exist from gravitational effects on visible matter, being undetectable by emitted or scattered electromagnetic radiation.  

Evidence from the formation of large-scale structure (galaxies and their clusters) strongly favour cosmologies where non-baryonic DM is entirely composed of cold dark matter (CDM), i.e.\ non-relativistic particles. CDM particles, in turn, may be weakly interacting massive particles (WIMPs), a class of DM candidates that arise naturally in models which attempt to explain the origin of electroweak symmetry breaking. Furthermore, the typical (weak-scale) cross sections characterising these models are of the same order of magnitude as the WIMP annihilation cross section, thus establishing the so-called \emph{WIMP miracle}; this is precisely where the connection between Cosmology and Particle Physics lies~\cite{mavromatos}. 

WIMP dark matter candidates include the lightest neutralino in models with weak-scale supersymmetry~\cite{susy-dm}, Kaluza-Klein photons arise in scenarios with universal extra dimensions (UED)~\cite{ued-dm}, while lightest $T$-odd particles are predicted in Little Higgs models~\cite{little} with a conserved $T$-parity. The common denominator in these theories is that they all predict the existence of an electrically neutral, colorless and \emph{stable} particle, whose decay is prevented by a kind of symmetry: \R-parity, connected to baryon and lepton number conservation in SUSY models; KK-parity, the four-dimensional remnant of momentum conservation in extra dimension scenarios; and a $Z_2$ discrete symmetry called $T$-parity in Little Higgs models. 

Weakly interacting massive particles do not interact neither electromagnetically nor strongly with matter and thus, once produced, they traverse the various detectors layers without leaving a trace, just like neutrinos. However by exploiting the hermeticity of the experiments, we can get a hint of the WIMP presence through the balance of the energy/momentum measured in the various detector components, the so-called \emph{missing energy}. In hadron colliders, in particular, since the longitudinal momenta of the colliding partons are unknown, only the \emph{transverse missing energy}, \met, can be reliably used to `detect' DM particles. 

\subsection{The ATLAS and CMS experiments at the LHC}\label{sc:lhc}

The Large Hadron Collider (LHC)~\cite{lhc}, situated at CERN, the European Laboratory for Particle Physics, outside Geneva, Switzerland, started its physics program in 2010 colliding two counter-rotating beams of protons or heavy ions. Before the scheduled 2013--2014 long shutdown, the LHC succeeded in delivering $\sim5~\ifb$ of integrated luminosity at centre-of-mass energy of $7~\tev$ during 2010--2011 and another $\sim23~\ifb$ at $\sqrt{s}=8~\tev$ in 2012. The LHC has already extended considerably the reach of its predecessor hadron machine, the Fermilab Tevatron, both in terms of instantaneous luminosity and energy, despite the fact that it has not arrived yet to its design capabilities. 

The two general-purpose experiments, ATLAS (A Toroidal LHC ApparatuS)~\cite{atlas-det} and CMS (Compact Muon Solenoid)~\cite{cms-det}, have been constructed and operate with the aim of exploring a wide range of possible signals of New Physics that LHC renders accessible, on one hand, and performing precision measurements of Standard Model (SM) parameters, on the other. It is worth mentioning that the MoEDAL~\cite{moedal} experiment is specifically designed to explore high-ionisation signatures that may also arise in some theoretical scenarios of dark matter~\cite{moedal-review}. 

The ATLAS~\cite{atlas-det} and CMS~\cite{cms-det} detectors are designed to overcome difficult experimental challenges: high radiation levels, large interaction rate and extremely small production cross sections of New Physics signals with respect to known SM processes. To this end, both experiments feature separate subsystems to measure charged particle momentum, energy deposited by electromagnetic showers from photons and electrons, energy from hadronic showers of strongly-interacting particles and muon-track momentum.

The most remarkable highlight of ATLAS and CMS operation so far is undoubtedly the discovery of a new particle~\cite{higgs-disc} that so far seems to have all the features pinpointing to a SM(-like) Higgs boson~\cite{higgs-prop}. The observation of this new boson has strong impact not only on our understanding of the fundamental interactions of Nature, as encoded in the SM, but on the proposed theoretical scenarios of Physics beyond the SM (BSM).  

\section{Model-independent DM production at the LHC}\label{sc:monox}

Collider searches for dark matter are highly complementary to direct~\cite{col-direct,col-di-in} and indirect~\cite{col-di-in,col-indirect} DM detection methods. The main advantage of collider searches is that they do not suffer from astrophysical uncertainties and that there is no lower limit to their sensitivity on DM masses. The leading generic diagrams responsible for DM production~\cite{tevatron1,tevatron2} at hadron colliders involve the pair-production of WIMPs plus the initial- or final-state radiation (ISR/FSR) of a gluon, photon or a weak gauge boson $Z,\,W$. The ISR/FSR particle is necessary to balance the two WIMPs' momentum, so that they are not produced back-to-back resulting in negligible \met. Therefore the search is based on selecting events high-\met\ events, due to the WIMPs, and a single jet, photon or boson candidate. 



The search results are interpreted in terms of a largely model-independent effective-field-theory framework, in which the interactions between a DM Dirac fermion \x\ and SM fermions $f$ are described by contact operators. Some of the possible operators are listed in Table~\ref{tb:operators}. In this framework, the interaction between SM and DM particles is determined by only two parameters, namely the DM-particle mass, \mx, and the suppression scale, $M_*$, which is related to the mediator mass and to its coupling to SM and DM particles. The derived limits are independent of the theory behind the WIMP (SUSY, extra dimensions, etc), yet it \emph{has} been assumed that other hypothetical particles are too heavy to be produced directly in $pp$ collisions. Henceforth, Dirac DM fermions are considered, however conclusions for Majorana fermions can also be drawn, since their production cross section only differs by a factor of two. 

\begin{table}[htb]
\caption{\label{tb:operators}Effective interaction operators of WIMP pair production considered in the mono-$X$  analyses, following the formalism of Ref.~\cite{tevatron1}.}
\begin{center}
\begin{tabular}{@{}cccc@{}} 
\br
Name & Initial state & Type & Operator \\ \mr
D1  & $qq$ & scalar       & $\frac{m_q}{M_*^3}\bar{\x}\x\bar{q}q$ \\
D5  & $qq$ & vector       & $\frac{1}{M_*^2}\bar{\x}\gamma^{\mu}\x\bar{q}\gamma_{\mu}q$ \\
D8  & $qq$ & axial-vector & $\frac{1}{M_*^2}\bar{\x}\gamma^{\mu}\gamma^5\x\bar{q}\gamma_{\mu}\gamma^{\mu}q$ \\
D9  & $qq$ & tensor       & $\frac{1}{M_*^2}\bar{\x}\sigma^{\mu\nu}\x\bar{q}\sigma_{\mu\nu}q$ \\
D11 & $gg$ & scalar       & $\frac{1}{4M_*^3}\bar{\x}\x\alpha_s(G^s_{\mu\nu})^2$\\ 
\br
\end{tabular} 
\end{center}
\end{table}

\subsection{Monojet searches}\label{sc:monojet}

Event topologies with a single high-\et\ jet and large \met, hereafter referred to as \emph{monojets}, constitute valuable probes of physics beyond the SM at the LHC. Both ATLAS~\cite{atlas-monojet} and CMS~\cite{cms-monojet} experiments have performed searches for an excess of monojet events over SM expectations in a wide range of signatures. The analyses outlined here use the full 2011 $pp$ LHC dataset at a centre-of-mass energy of $\sqrt{s} = 7~\tev$. The primary SM process yielding a true monojet final state is $Z$-boson production in association with a jet, where the $Z$ decays to two neutrinos. Other known processes acting as background in this search are $Z(\to\ell\ell)$+jets ---with $\ell=e,\,\mu$---, \W+jets, $t\bar{t}$ as well as single-top events. All electroweak backgrounds and multijet events passing the selections criteria, as well as non-collision backgrounds, are in most cases determined by data-driven methods. Top and diboson backgrounds are estimated solely from Monte Carlo (MC) simulation.  

The monojet analyses for ATLAS and CMS are based on some general requirements: large \met, with thresholds typically ranging from 120~\gev\ to 500~\gev\ and a energetic jet with a variable \pt\ threshold higher than 110~\gev\ that fulfils high jet-reconstruction quality criteria. Moreover, events with at least one electron or muon or a third jet are vetoed. Back-to-back dijet events are suppressed by requiring the subleading jet not to be collinear with $\bf p_{\bf T}^{\bf miss}$. The selected data are required to pass a trigger based on high \met\ (ATLAS) or large \met\ plus one high-\et\ jet (CMS).  
   
The data, amounting to $\sim5~\ifb$, are found to be in agreement with the SM expectations. The results are interpreted in a framework of WIMP production with the simulated WIMP-signal MC samples corresponding to various assumptions of the effective field theory, as discussed previously. In this framework, the interaction between SM and DM particles are defined by only two parameters, namely the DM-particle mass, \mx, and the suppression scale, $M_*$, which is related to the mediator mass and to its coupling to SM and DM particles.

Experimental and theoretical systematic uncertainties are considered when setting limits on the model parameters $M_*$ and \mx. The experimental uncertainties on jet energy scale and resolution and on \met\ range from $1-20\%$ of the WIMP event yield, depending on the \met\ and \pt\ thresholds and the considered interaction operator. Other experimental uncertainties include the ones associated with the trigger efficiency and the luminosity measurement. On the other hand, the parton-distribution-function set, the amount of ISR/FSR, and the factorisation and renormalisation scales assumed lead to theoretical uncertainties on the simulated WIMP signal. 

\begin{figure}[htb]
\centerline{\includegraphics[width=0.48\textwidth]{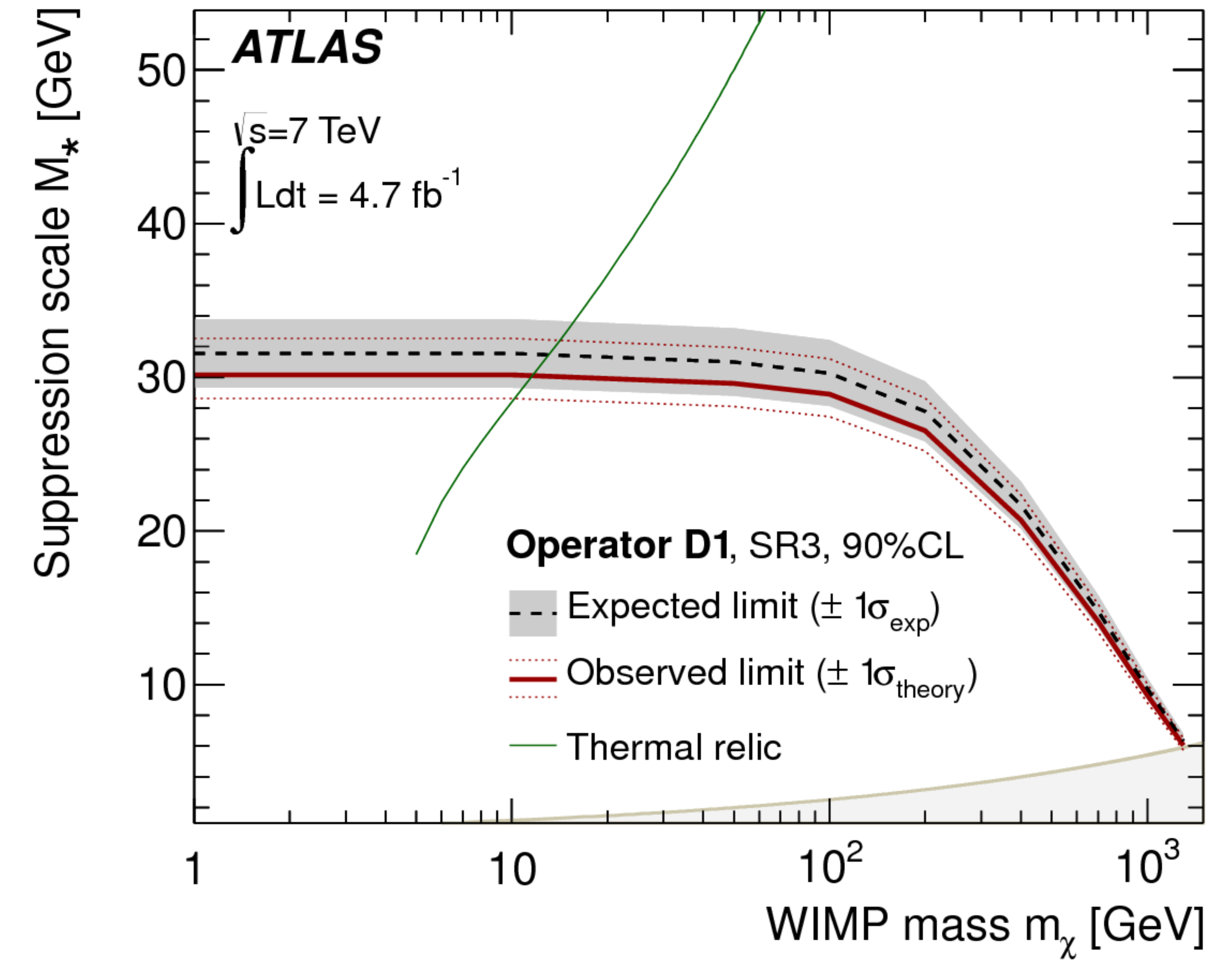}
  \hspace{0.02\textwidth}
  \includegraphics[width=0.48\textwidth]{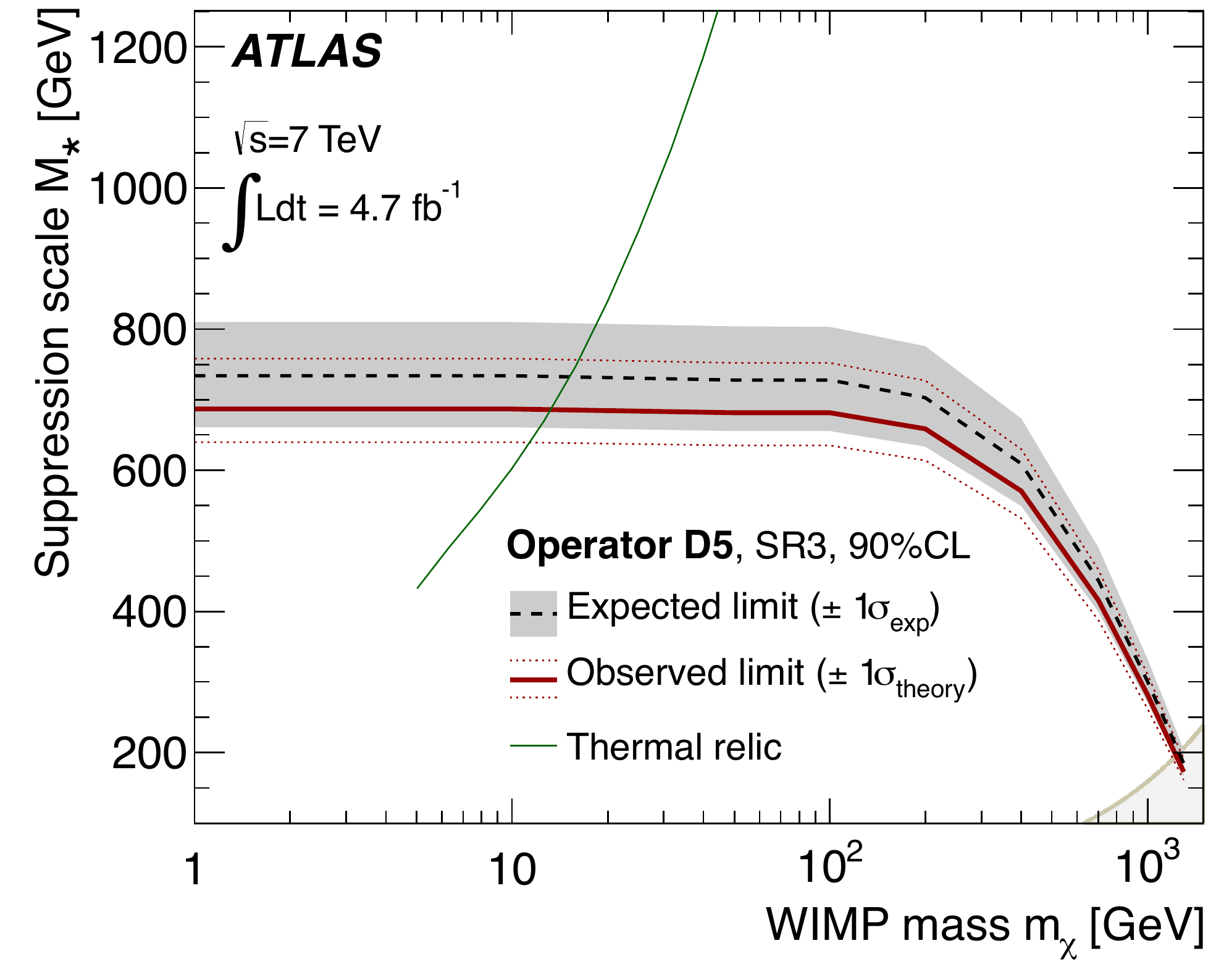}}
\centerline{\includegraphics[width=0.48\textwidth]{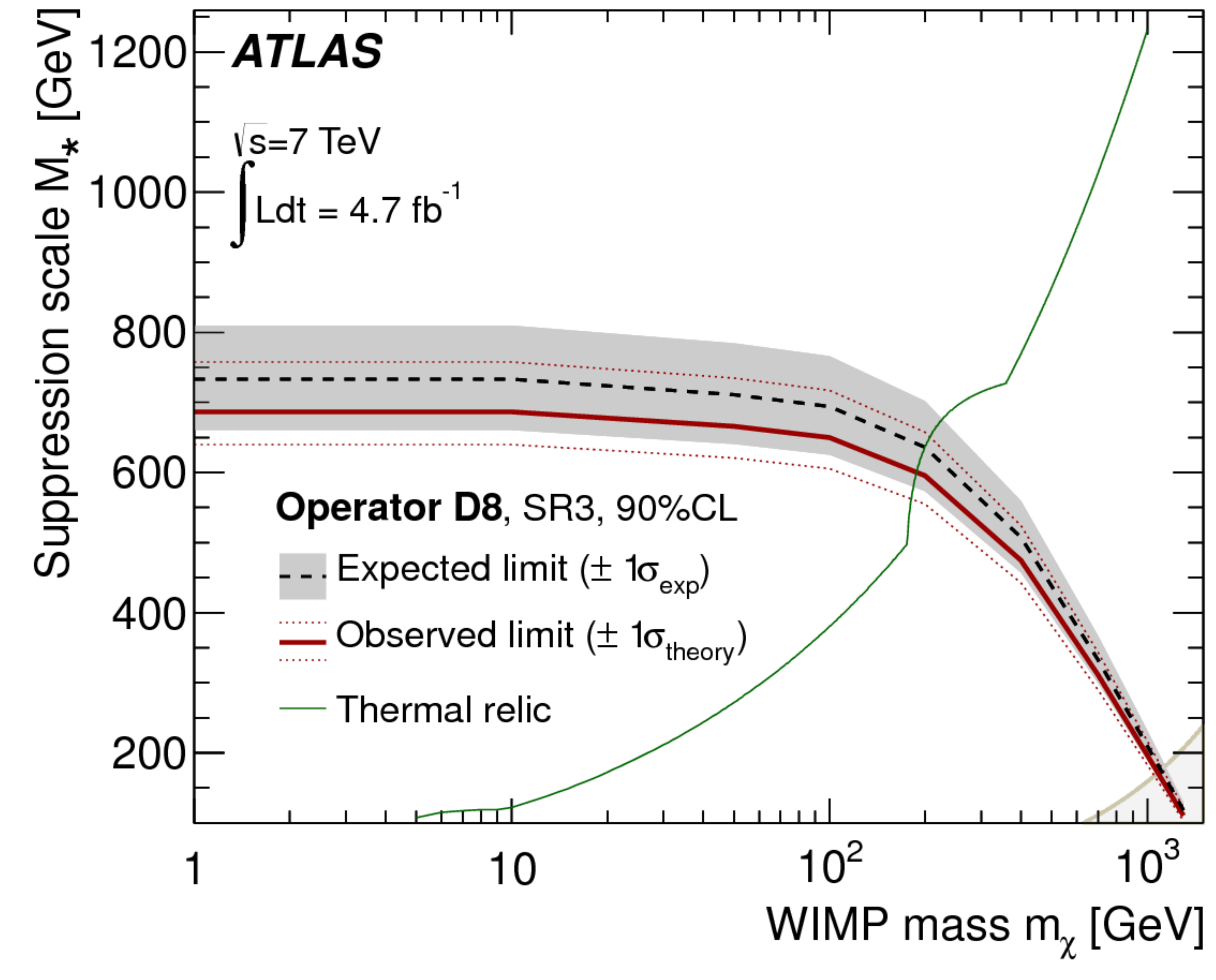}} 
\caption{ATLAS lower limits at 90\% CL on the suppression scale, $M_*$, for different masses of \x\ obtained with the monojet analysis for the operators D1 (top left), D5 (top right) and D8 (bottom). The region below the limit lines is excluded. All shown curves and areas are explained in the text. From Ref.~\cite{atlas-monojet}.\label{fg:atlas-monojet-d}}
\end{figure}

From the limit on the visible cross section of new BSM physics processes, lower limits on the suppression scale as a function of the WIMP mass have been derived by the ATLAS Collaboration~\cite{atlas-monojet}. The 90\% confidence level (CL) lower limits for the D1, D5 and D8 operators are shown in Fig.~\ref{fg:atlas-monojet-d}. The observed limit on $M_*$ includes experimental uncertainties; the effect of theoretical uncertainties is indicated by dotted $\pm1\sigma$ lines above and below it. Around the expected limit, $\pm1\sigma$ variations due to statistical and systematic uncertainties are shown as a grey band. The lower limits are flat up to $\mx\lesssim100~\gev$ and become weaker at higher mass due to the collision energy. In the bottom-right corner of the $(\mx,\,M_*)$ plane (light-grey shaded area), the effective field theory approach is no longer valid. The rising lines correspond to couplings consistent with the measured thermal relic density~\cite{tevatron1}, assuming annihilation in the early universe proceeded exclusively via the given operator. Similar exclusion limits for all operators listed in Table~\ref{tb:operators} are given in Ref.~\cite{atlas-monojet}. For the operator D1, the limits are much weaker ($\sim30~\gev$) than for other operators. Nevertheless, if heavy-quark loops are included in the analysis, much stronger bounds on $M_*$ can be obtained~\cite{monojet-loop}.

The observed limit on the dark matter-nucleon scattering cross section depends on the mass of the dark matter particle and the nature of its interaction with the SM particles. The limits on the suppression scale as a function of \mx\ can be translated into a limit on the cross section using the reduced mass of \x-nucleon system~\cite{tevatron2}, which can be compared with the constraints from direct and indirect detection experiments, as we shall see at the end of this Section in conjunction with bounds acquired with other mono-$X$ analyses.

\begin{figure}[htb]
\includegraphics[width=0.5\textwidth]{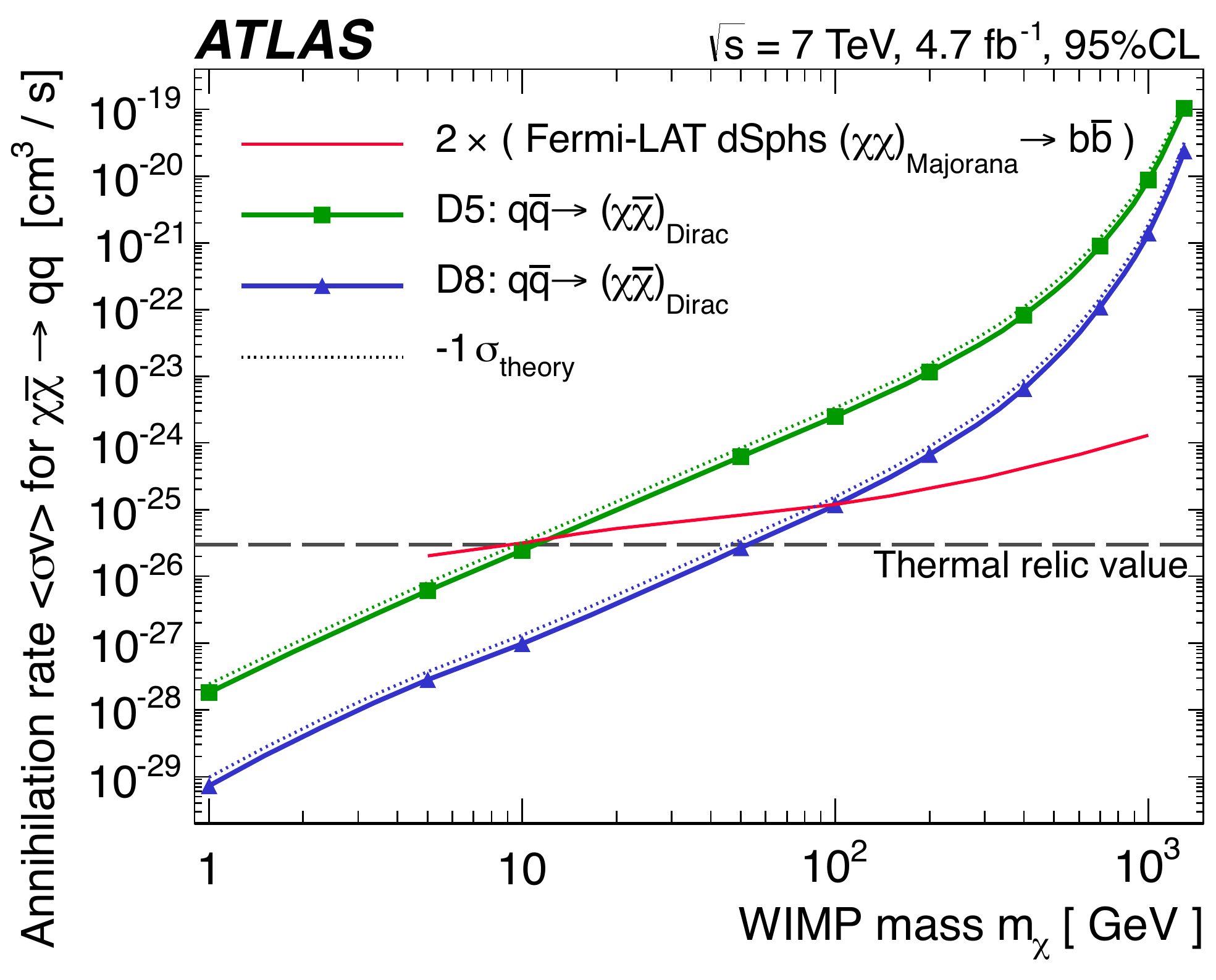}\hspace{2pc}%
\begin{minipage}[b]{0.4\textwidth}\caption{\label{fg:atlas-monojet-annih}ATLAS 95\% CL limits on WIMP annihilation rates $\langle\sigma v\rangle$ versus mass \mx, inferred from the monojet analysis. Explanation of the shown curves is given in the text. From Ref.~\cite{atlas-monojet}.}
\end{minipage}
\end{figure}

The ATLAS collider limits on vector (D5) and axial-vector (D8) interactions are also interpreted in terms of the relic abundance of WIMPs, using the same effective theory approach~\cite{tevatron1}. The upper limits on the annihilation rate of WIMPs into light quarks, defined as the product of the annihilation cross section $\sigma$ and the relative WIMP velocity $v$ averaged over the WIMP velocity distribution $\langle\sigma v\rangle$, are shown in Fig.~\ref{fg:atlas-monojet-annih}. The results are compared to limits on WIMP annihilation to $b\bar{b}$, obtained from galactic high-energy gamma-ray observations, measured by the Fermi-LAT telescope~\cite{fermilat}. Gamma-ray spectra and yields from WIMPs annihilating to $b\bar{b}$, where photons are produced in the hadronisation of the quarks, are expected to be very similar to those from WIMPs annihilating to light quarks~\cite{annihilation}. Under this assumption, the ATLAS and Fermi-LAT limits can be compared, after scaling up the Fermi-LAT values by a factor of two to account for the Majorana-to-Dirac fermion adaptation. Again, the ATLAS bounds are especially important for small WIMP masses: below 10~\gev\ for vector couplings and below about $100~\gev$ for axial-vector ones. In this region, the ATLAS limits are below the annihilation cross section needed to be consistent with the thermic relic value, keeping the assumption that WIMPs have annihilated to SM quarks only via the particular operator in question. For masses of $\mx\gtrsim 200~\gev$ the ATLAS sensitivity becomes worse than the Fermi-LAT one. In this region, improvements can be expected when going to larger centre-of-mass energies at the LHC.

The case in which the mediator is light enough to be accessible to the LHC has been considered too by the CMS experiment in a monojet search performed with $\sim20~\ifb$ at 8~\tev~\cite{cms-monojet-lm}. Figure~\ref{fg:cms-light-mediator} shows the observed limits on the contact interaction scale $\Lambda$ as a function of the mass of the mediator $M$, assuming vector interactions and a dark matter mass of 50~\gev\ and 500~\gev. The width  $\Gamma$ of the mediator is varied between $M/3$ and $M/8\pi$~\cite{fox}. It shows the resonant enhancement in the production cross section once the mass of the mediator is within the kinematic range and can be produced on-shell. For $m_{\x}\gtrsim 100~\gev$, this approach is adequate and quite conservative in the bounds on $\Lambda$. For $m_{\x}\lesssim 100~\gev$, the collider bounds are considerably weaker. At large mediator masses, i.e.\ $M\gtrsim 5~\tev$, the limits on $\Lambda$ approximate to those obtained in the effective theory framework. 

\begin{figure}[htb]
\includegraphics[width=0.5\textwidth]{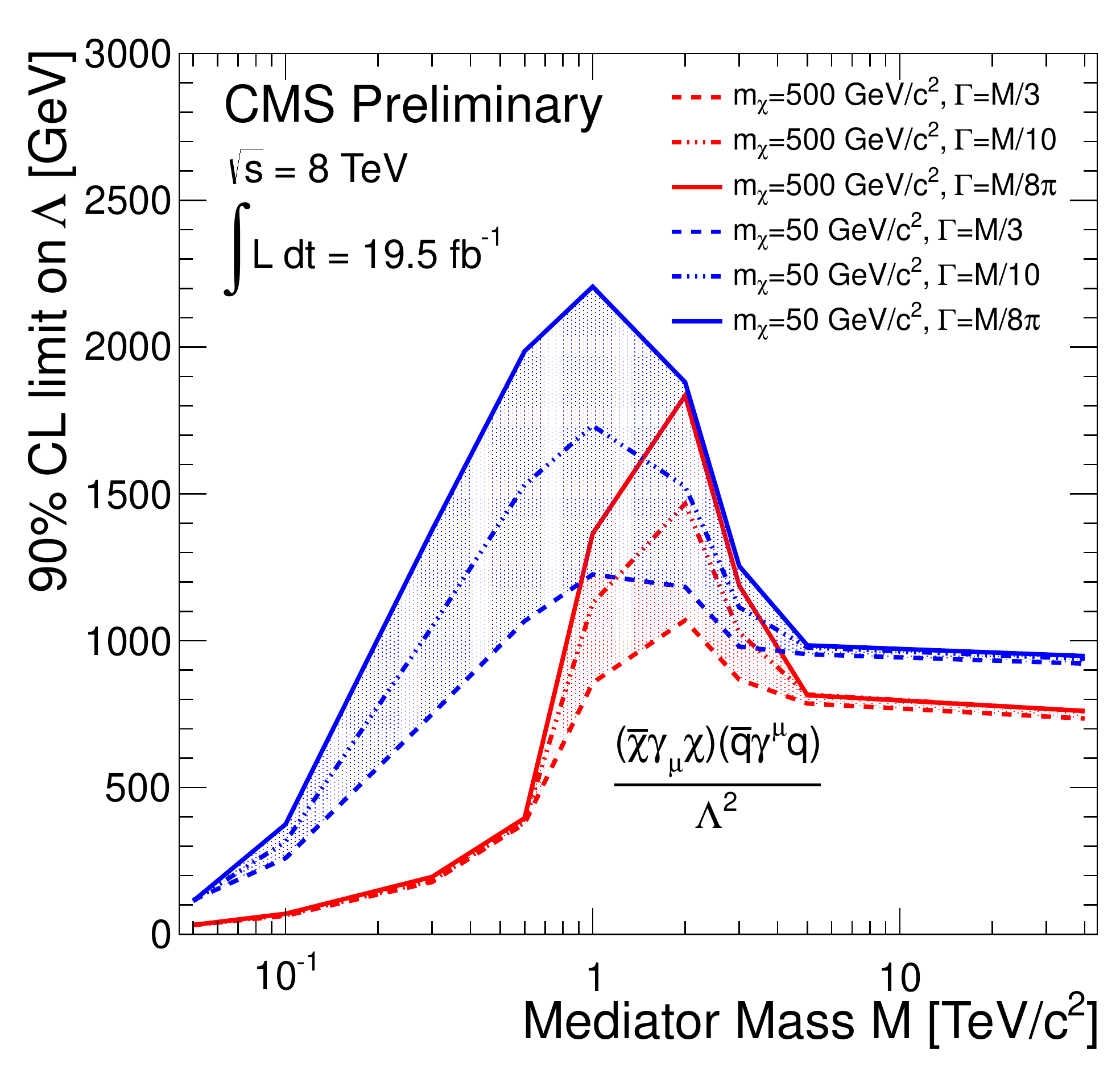}\hspace{2pc}%
\begin{minipage}[b]{0.4\textwidth}\caption{\label{fg:cms-light-mediator}Observed limits on the scale $\Lambda$ as a function of the mass $M$, assuming vector interactions and a dark matter mass of $50~\gev$ (blue) and $500~\gev$ (red) in a CMS monojet analysis. The width of the mediator was varied between $M/3$, $M/10$ and $M/8\pi$. From Ref.~\cite{cms-monojet-lm}.}
\end{minipage}
\end{figure}

\subsection{Monophoton-based probes}\label{sc:monophoton}

In the same fashion as in the monojet searches, the \emph{monophoton} analyses aim at probing dark matter requiring large \met\ ---from the \x-pair production--- and at least one ISR/FSR photon. Searches in monophoton events by ATLAS~\cite{atlas-monophoton} and CMS~\cite{cms-monophoton} also show an agreement with the SM expectations. The limits are derived in a similar fashion as for the monojet search, however the monophoton search is found to be somewhat less sensitive with respect to the monojet topology.

The primary (irreducible) background for a $\gamma+\met$ signal comes from $Z\gamma\to\nu\bar{\nu}\gamma$ production. This together with other SM backgrounds, including $W\gamma$, $W\to e\nu$, $\gamma+\text{jet}$ multijet, diphoton and diboson events, as well as backgrounds from beam halo and cosmic-ray muons, are taken into account in the analyses. The CMS analysis is based on singe-photon triggers, whilst ATLAS relies on high-\met\ triggered events. The photon candidate is required to pass tight quality and isolation criteria, in particular in order to reject events with electrons faking photons. The missing transverse momentum of the selected events should be as high as 150~\gev\ (130~\gev) in the ATLAS (CMS) search. In CMS, events with a reconstructed jet are vetoed, while the ATLAS analysis rejects events with an electron, a muon or a second jet. 

Both analyses, observe no significant excess of events over the expected background when applied on $\sim5~\ifb$ of $pp$ collision data at $\sqrt{s}=7~\tev$. Hence they set lower limits on the suppression scale, $M_*$ versus the DM fermion mass, \mx, which in turn they are translated into upper limits on the nucleon-WIMP interaction cross section applying the prescription in Ref.~\cite{tevatron1}. Figure~\ref{fg:atlas-monophoton} shows the 90\% CL upper limits on the nucleon-WIMP cross section as a function of \mx\ derived from the ATLAS search~\cite{atlas-monophoton}. The results are compared with previous CDF~\cite{cdf2}, CMS~\cite{cms-monojet,cms-monophoton} and direct WIMP detection experiments~\cite{xenon100a,cdmsii,cogent,simple,picasso2} results. The CMS limit curve generally overlaps the ATLAS curve.

\begin{figure}[htb]
\centerline{\includegraphics[width=0.85\textwidth]{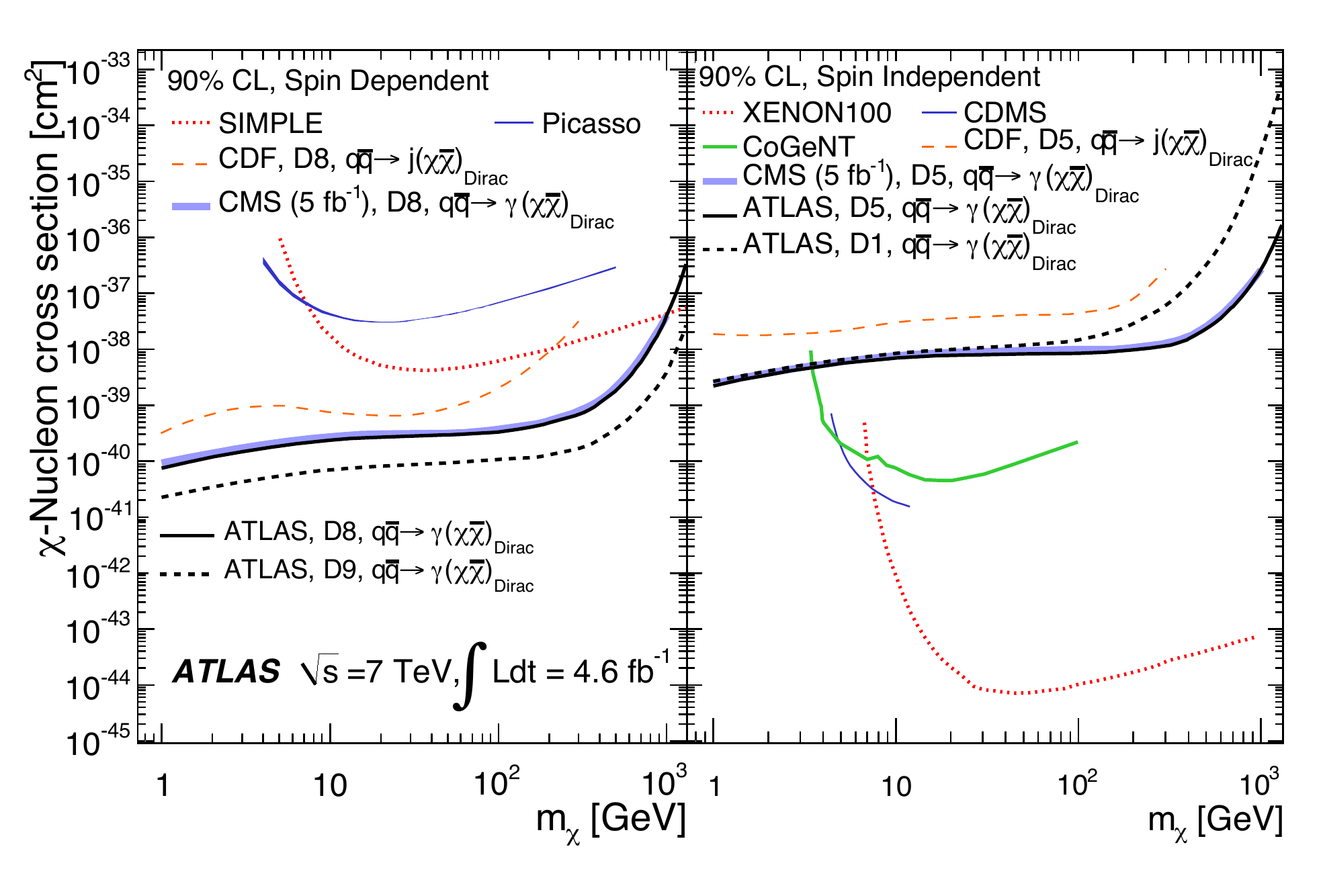}}
\caption{ATLAS 90\% CL upper limits on the nucleon-WIMP cross section as a function of \mx\ for spin-dependent (left) and spin-independent (right) interactions, corresponding to D8, D9, D1, and D5 operators, derived from the monophoton analysis. Explanation of the shown curves is given in the text. From Ref.~\cite{atlas-monophoton}.\label{fg:atlas-monophoton}}
\end{figure}

The observed limits on $M_*$ typically decrease by 2\% to 10\% if the $-1\sigma$ theoretical uncertainty, resulting from the same sources as the one cited in the monojet analysis, is considered. This translates into a 10\% to 50\% increase of the quoted nucleon-WIMP cross section limits. To recapitulate, the exclusion in the region $1~\gev<\mx<1~\tev$ ($1~\gev<\mx<3.5~\gev$) for spin-dependent (spin-independent) nucleon-WIMP interactions is driven by the results from collider experiments, always under the assumption of the validity of the effective theory, and is still dominated by the monojet results.

\subsection{Mono-$W$ and mono-$Z$ final states}\label{sc:monoWZ}

As demonstrated in the previous sections, searches for monojet or monophoton signatures have yielded powerful constraints on dark matter interactions with SM particles. Other studies propose probing DM at LHC through a  
$pp\to\x\bar{\x}+W/Z$, with a leptonically decaying $W$~\cite{monow} or $Z$~\cite{monoz}. The final state in this case would be large \met\ and a single charged lepton (electron or muon) for the \emph{mono-W} signature (\emph{monolepton}) or a pair of charged leptons that reconstruct to the $Z$ mass for the \emph{mono-Z} signature. In either case, the gauge boson radiations off a $q\bar{q}$ initial state and an effective field theory is deployed to describe the  contact interactions that couple the SM particle with the WIMP.  

In Ref.~\cite{cms-monow}, the existing $W'$ searches from CMS~\cite{cms-wprime8} ---which share a similar final state with mono-$W$ searches--- are used to place a bound on mono-$W$ production at LHC, which for some choices of couplings are better than monojet bounds. This is illustrated in the left (right) panel of Fig.~\ref{fg:cms-monolep}, where the spin-dependent (spin-independent) WIMP-proton cross section limits are drawn. The parameter $\xi$ parametrises the relative strength of the coupling to down-quarks with respect to up-quarks: $\xi=+1$ for equal couplings; $\xi=-1$ for opposite-sign ones; and $\xi=0$ when there is no coupling to down-quarks. Even in cases where the monoleptons do not provide the most stringent constraints, they provide an interesting mechanism to disentangle WIMP couplings to up-type versus down-type quarks. Such an interpretation has also been performed in Ref.~\cite{monow} yielding similar limits.  

\begin{figure}[htb]
\centerline{\includegraphics[width=0.48\textwidth]{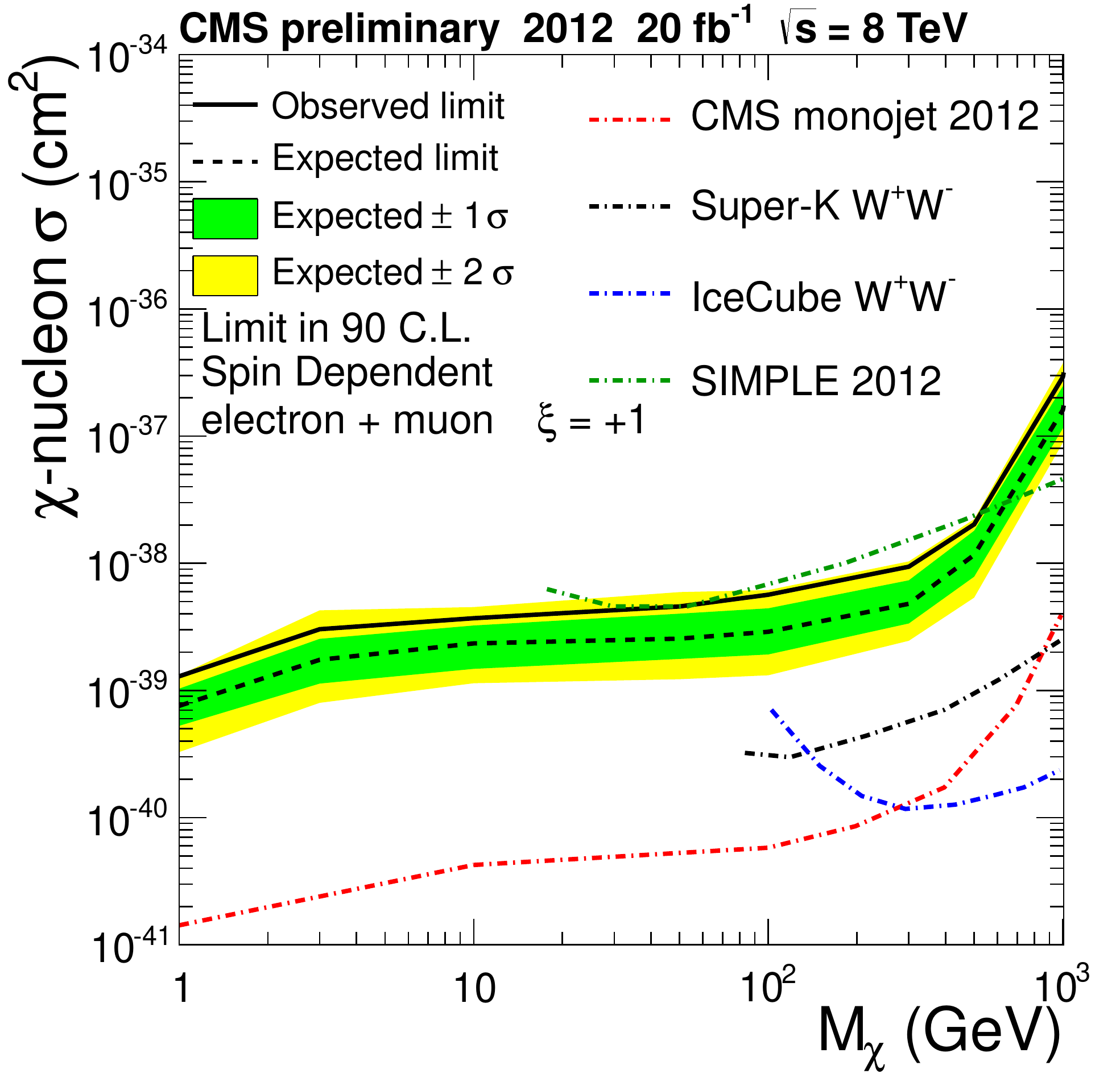}
  \hspace{0.02\textwidth}
  \includegraphics[width=0.48\textwidth]{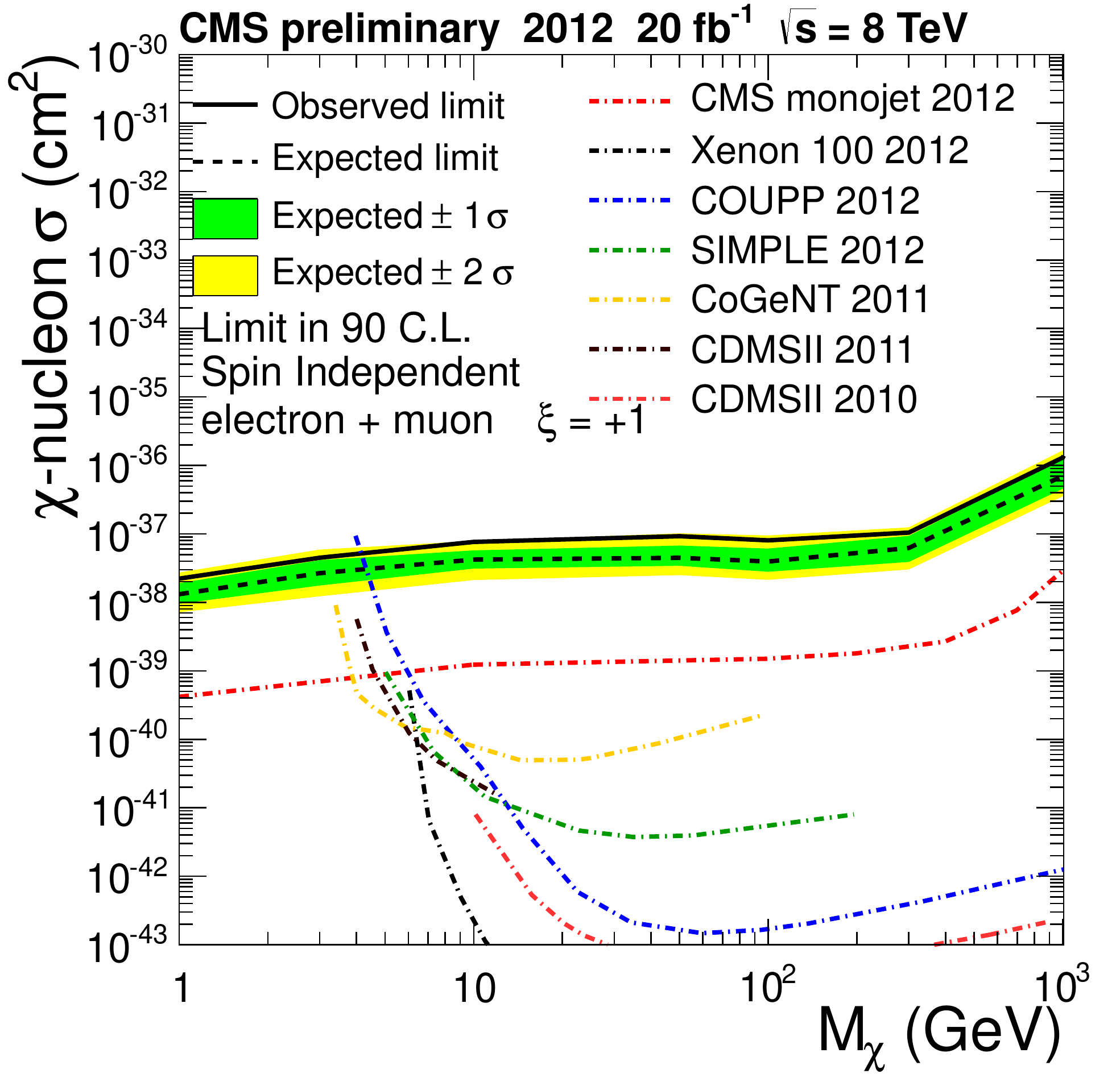}}
\caption{CMS monolepton search with 20 ~\ifb\ at 8~\tev. Excluded proton-dark matter cross section for axial-vector-like, i.e.\ spin dependent (left), and vector-like, i.e.\ spin independent (right), for the combination of electron and muon channels. The CMS monojet result is for 20~\ifb\ of 2012 data~\cite{cms-monojet-lm}. From Ref.~\cite{cms-monow}.\label{fg:cms-monolep}}
\end{figure}
 
The ATLAS Collaboration has extended the range of possible mono-$X$ probes by looking for $pp\to\x\bar{\x}+W/Z$, when the gauge boson decays to two quarks~\cite{atlas-monowz}, as opposed to the leptonic signatures discussed so far. The analysis searches for the production of $W$ or $Z$ bosons decaying hadronically and reconstructed as a single massive jet in association with large \met\ from the undetected $\x\bar{\x}$ particles. For this analysis, the jet candidates are reconstructed using a filtering procedure referred to as \emph{large-radius jets}~\cite{fat-jets}. This search, the first of its kind, is sensitive to WIMP pair production, as well as to other DM-related models, such as invisible Higgs boson decays ($WH$ or $ZH$ production with $H\to\x\bar{\x}$).

\begin{figure}[htb]
\centerline{\includegraphics[width=0.8\textwidth]{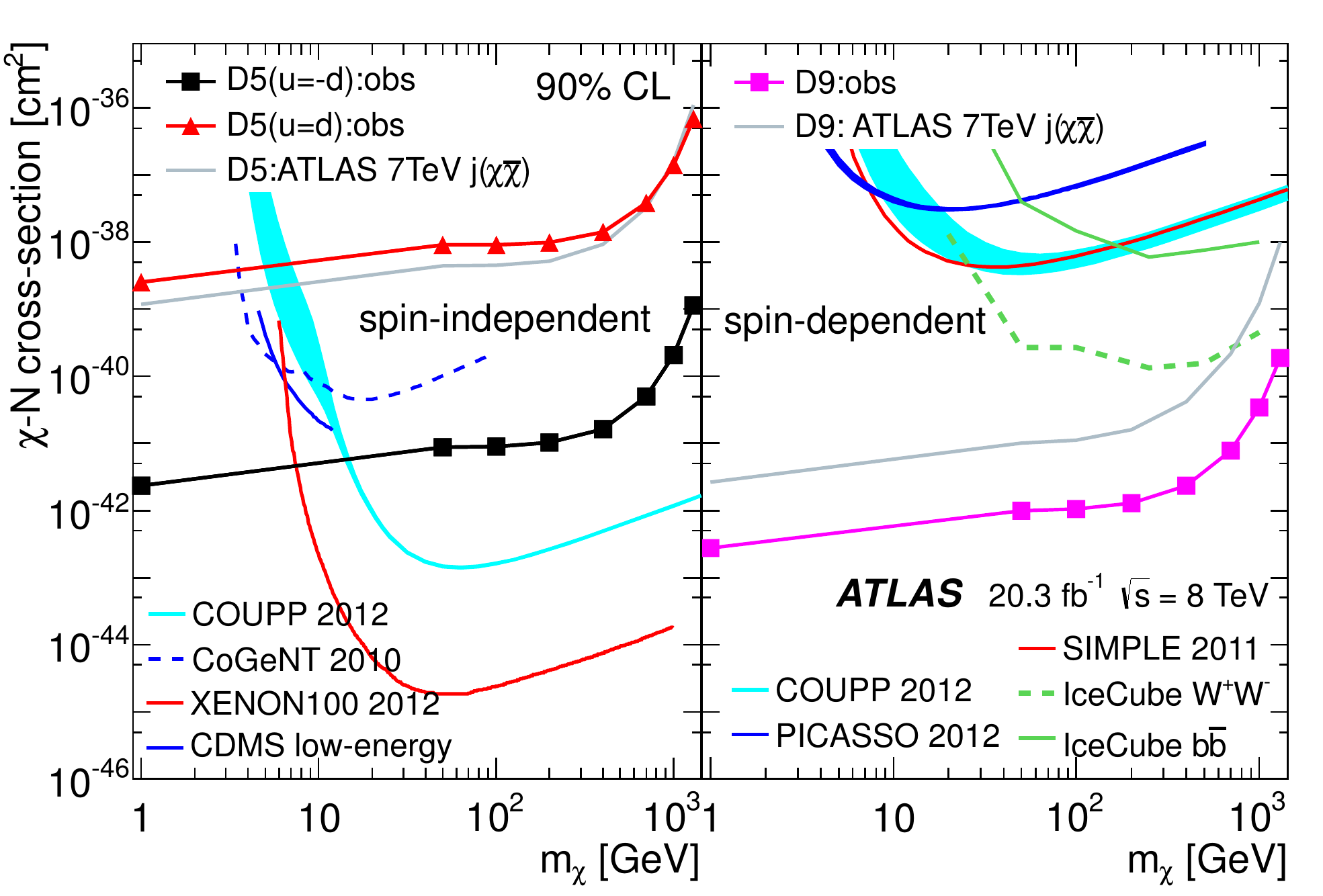}}
\caption{ATLAS-derived limits on \x-nucleon cross sections as a function of \mx\ at 90\% CL for spin-independent (left) and spin-dependent (right) cases, obtained with the mono-$W/Z$ analysis and compared to previous limits. From Ref.~\cite{atlas-monowz}.\label{fg:atlas-monowz}}
\end{figure}

Figure~\ref{fg:atlas-monowz} shows the 90\% CL upper limits on the dark matter-nucleon scattering cross section as a function of the mass of DM particle for the spin-independent (left) and spin-dependent (right) models obtained by the ATLAS mono-$W/Z$ analysis~\cite{atlas-monowz}. The new limits are also compared to the limits set by ATLAS in the $7~\tev$ monojet analysis~\cite{atlas-monojet}. Limits from XENON100~\cite{xenon100b}, CoGent~\cite{cogent}, CDMS~II~\cite{cdmsii}, SIMPLE~\cite{simple}, COUPP~\cite{coupp2}, Picasso~\cite{picasso2}, IceCube~\cite{icecube2} are superimposed for comparison. For the spin-independent case with the opposite-sign up-type and down-type couplings, the limits are improved by about three orders of magnitude. For other cases, the bounds are similar. Comparable limits have been obtained by the CMS experiment.

It is worth noting that the spin-dependent limits derived from the operator D9, give a smaller, hence better, bound on the WIMP-nucleon cross section throughout the range of \mx, compared to direct DM experiments. In the spin-independent case the bounds from direct detection experiments are stronger for $\mx\gtrsim10~\gev$, whereas the collider bounds, acquired with the operator D5, get important for the region of low DM masses. 

\section{Searches for supersymmetry}\label{sc:susy}

Supersymmetry (SUSY)~\cite{susy} is an extension of the Standard Model which assigns to each SM field a superpartner field with a spin differing by a half unit. SUSY provides elegant solutions to several open issues in the SM, such as the hierarchy problem and the grand unification. In particular, SUSY predicts the existence of a stable weakly interacting particle ---the lightest supersymmetric particle (LSP)--- that has the pertinent properties to be a dark matter particle, thus providing a very compelling argument in favour of SUSY~\cite{lisboa}.

SUSY searches in the ATLAS~\cite{atlas-det} and CMS~\cite{cms-det} experiments typically focus on events with high transverse missing energy, which can arise from (weakly interacting) LSPs, in the case of \R-parity conserving SUSY, or from neutrinos produced in LSP decays, if \R-parity is broken (c.f.\ Section~\ref{sc:rpv}). Hence, the event selection criteria of inclusive channels are based on large \met, no or few leptons ($e$, $\mu$), many jets and/or $b$-jets, $\tau$-leptons and photons. In addition, kinematical variables such as the transverse mass, \mt, and the effective mass, \mef, assist in distinguishing further SUSY from SM events, whilst the \emph{effective transverse energy}~\cite{alberto-ete} can be useful to cross-check results, allowing a better and more robust identification of the SUSY mass scale, should a positive signal is found. Although the majority of the analysis simply look for an excess of events over the SM background, there is an increasing application of distribution shape fitting techniques~\cite{shape}.  

Typical SM backgrounds are top-quark production  ---including single-top---, $W$/$Z$ in association with jets, dibosons and QCD multijet events. These are estimated using semi- or fully data-driven techniques. Although the various analyses are optimised for a specific SUSY scenario, the interpretation of the results are extended to various SUSY models or topologies. 

Analyses exploring $R$-parity conserving SUSY models at LHC are roughly divided into inclusive searches for squarks and gluinos, for third-generation fermions, and for electroweak production (pairs of $\tilde{\chi}^0$, $\tilde{\chi}^{\pm}$, $\tilde{\ell}$). Although these searches are designed and optimised to look for $R$-parity conserving SUSY, interpretation in terms of $R$-parity violating (RPV) models is also possible. Other analyses are purely motivated by oriented by RPV scenarios and/or by the expectation of long-lived sparticles. Recent summary results from each category of ATLAS and CMS searches are presented in this section. 

\subsection{Gluinos and first two generations of quarks}\label{sc:strong}

At the LHC, supersymmetric particles are expected to be predominantly produced hadronically, i.e.\ through gluino-pair, squark-pair and squark-gluino production. Each of these (heavy) sparticles is going to decay into lighter ones in a cascade decay that finally leads to an LSP, which in most of the scenarios considered is the lightest neutralino \X. The two LSPs would escape detection giving rise to high transverse missing energy, hence the search strategy followed is based on the detection of high \met, many jets and possibly energetic leptons. The analyses make extensive use of data-driven Standard Model background measurements. 

The most powerful of the existing searches are based on all-hadronic final states with large missing transverse momentum~\cite{atlas-0l,cms-0l}. In the 0-lepton search, events are selected based on a jet+\met\ trigger, applying a lepton veto, requiring a minimum number of jets, high \met, and large azimuthal separation between the \met\ and reconstructed jets, in order to reject multijet background. In addition, searches for squark and gluino production in a final state with one or two leptons have been performed~\cite{atlas-lep,cms-lep}. The events are categorised by whether the leptons have higher or lower momentum and are referred to as the \emph{hard} and \emph{soft} lepton channels respectively. The soft-lepton analysis which enhances the sensitivity of the search in the difficult kinematic region where the neutralino and gluino masses are close to each other forming the so-called \emph{compressed spectrum.}~\cite{compressed} Leptons in the soft category are characterised by low lepton-\pt\ thresholds ($6-10~\gev$) and such events are triggered by sufficient \met. Hard leptons pass a threshold of $\sim25~\gev$ and are seeded with both lepton and \met\ triggers. Analyses based on the \emph{razor}~\cite{razor} variable have also been carried out by both experiments~\cite{atlas-razor,cms-razor}.

The major backgrounds ($t\bar{t}$, $W$+jets, $Z$+jets) are estimated by isolating each of them in a dedicated control region, normalising the simulation to data in that control region, and then using the simulation to extrapolate the background expectations into the signal region. The multijet background is determined from the data by a matrix method. All other (smaller) backgrounds are estimated entirely from the simulation, using the most accurate theoretical cross sections available. To account for the cross-contamination of physics processes across control regions, the final estimate of the background is obtained with a simultaneous, combined fit to all control regions.

\begin{figure}[htb]
\centerline{\includegraphics[width=0.8\textwidth]{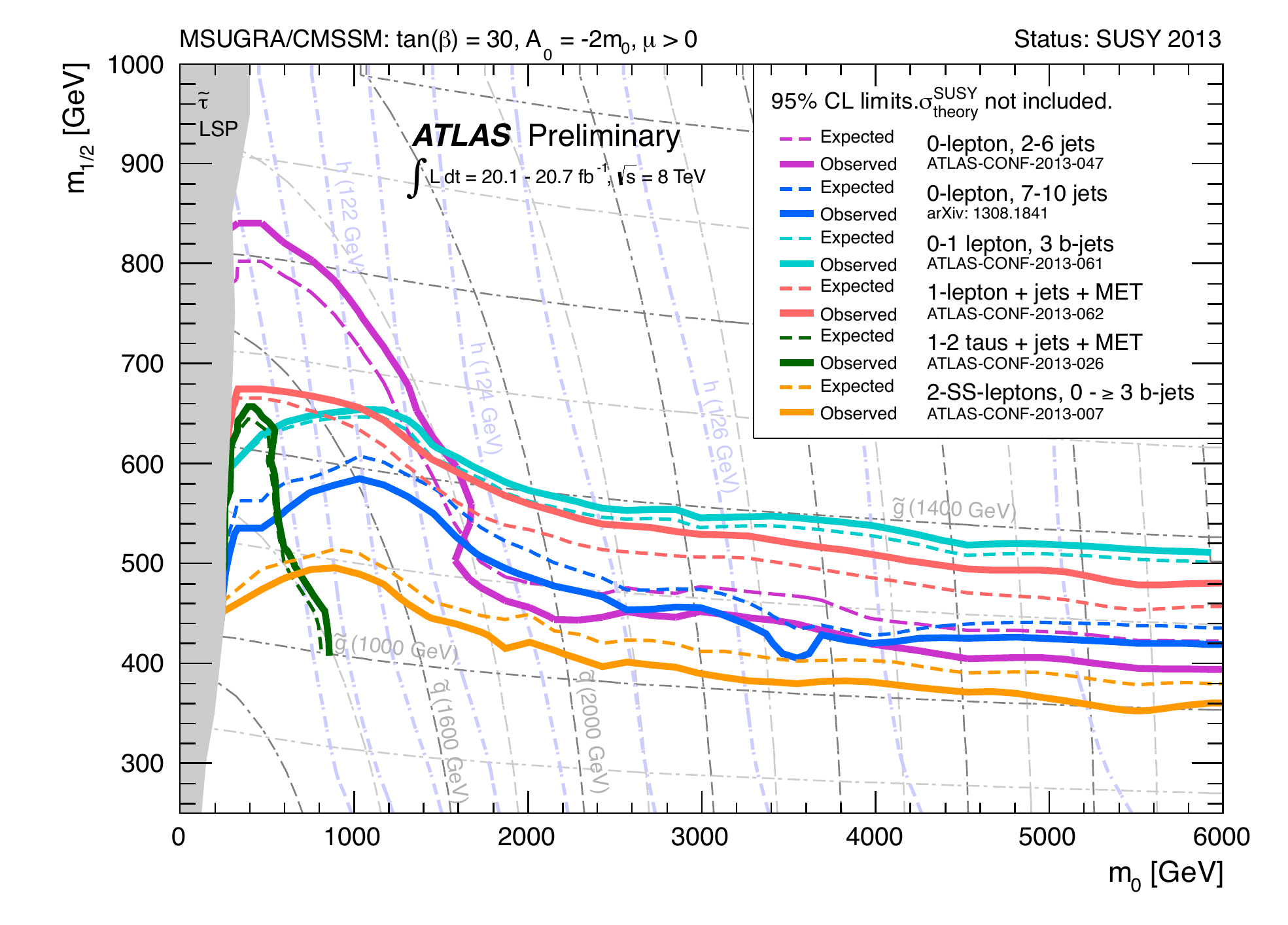}}
\caption{\label{fg:msugra}Exclusion limits at 95\% CL for 8~\tev\ ATLAS analyses~\cite{atlas-0l,atlas-msugra} in the $(m_0,\,m_{1/2})$ plane for the mSUGRA model. From Ref.~\cite{atlas-susy-results}.}
\end{figure}

In the absence of deviations from SM predictions, limits for squark and gluino production are set. Figure~\ref{fg:msugra} illustrates the 95\% CL limits set by ATLAS under the minimal Supergravity (mSUGRA) model in the $(m_0,\,m_{1/2})$ plane~\cite{atlas-0l,atlas-msugra}. The remaining parameters are set to $\tan\beta = 30$, $A_0 = -2\,m_0$, $\mu > 0$, so as to acquire parameter-space points where the predicted mass of the lightest Higgs boson, $h^0$, is near $125~\gev$, i.e.\ compatible with the recently observed Higgs-like boson~\cite{higgs-disc,higgs-prop}. Exclusion limits are obtained by using the signal region with the best expected sensitivity at each point. By assumption, the mSUGRA model avoids both flavour-changing neutral currents and extra sources of $CP$ violation. For masses in the TeV range, it typically predicts too much cold dark matter, however these predictions depend of the presence of stringy effects that may dilute~\cite{mavro} or enhance~\cite{vergou} the predicted relic dark matter density. In the mSUGRA case, the limit on squark mass reaches 1750~\gev\ and on gluino mass is 1400~\gev. 

\subsection{Third-generation squarks}\label{sc:third}

The previously presented limits from inclusive channels indicate that the masses of gluinos and first/second generation squarks are expected to be above 1~\tev. Nevertheless, in order to solve the hierarchy problem in a \emph{natural} way, the masses of the stops, sbottoms, higgsinos and gluinos have to be below the TeV-scale to properly cancel the divergences in the Higgs mass radiative corrections. Despite their production cross sections being smaller than for the first and second generation squarks, stop and sbottom may well be directly produced at the LHC and could provide the only direct observation of SUSY at the LHC in case the other sparticles are outside of the LHC energy reach. The lightest mass eigenstates of the sbottom and stop particles, $\t{b}_1$ and $\t{t}_1$, could hence be produced either directly in pairs or through $\t{g}$ pair production followed by $\t{g}\to\t{b}_1b$ or $\t{g}\to\t{t}_1t$ decays. Both cases will be discussed in the following.

For the aforementioned reasons, direct searches for third generation squarks have become a priority in both ATLAS and CMS. Such events are characterised by several energetic jets (some of them $b$-jets), possibly accompanied by light leptons, as well as high \met. A suite of channels have been considered, depending on the topologies allowed and the exclusions generally come with some assumptions driven by the shortcomings of the techniques and variables used, such as the requirement of 100\% branching ratios into particular decay modes.

In the case of the gluino-mediated production of stops, a simplified scenario (``Gtt model''), where $\tilde{t}_1$ is the lightest squark but $m_{\tilde{g}} < m_{\tilde{t}_1}$, has been considered. Pair production of gluinos is the only process taken into account since the mass of all other sparticles apart from the $\tilde{\chi}_1^0$ are above the \tev\ scale. A three-body decay via off-shell stop is assumed for the gluino, yielding a 100\% branching ratio for the decay $\tilde{g}\rightarrow t\bar{t}\tilde{\chi}_1^0$. The stop mass has no impact on the kinematics of the decay and the exclusion limits~\cite{cms-razor,cms-gluino,cms-ss2l} set by the CMS experiment are presented in the $(m_{\tilde{g}},m_{\tilde{\chi}_1^0})$ plane in Fig.~\ref{fg:gl-ttlsp}. For a massless LSP, gluinos with masses from 560~\gev\ to 1320~\gev\ are excluded. Similar results are obtained if the decay $\tilde{g}\rightarrow b\bar{b}\tilde{\chi}_1^0$ is considered instead, as shown in Fig.~\ref{fg:gl-bblsp}. 

\begin{figure}[htb]
\includegraphics[width=0.5\textwidth]{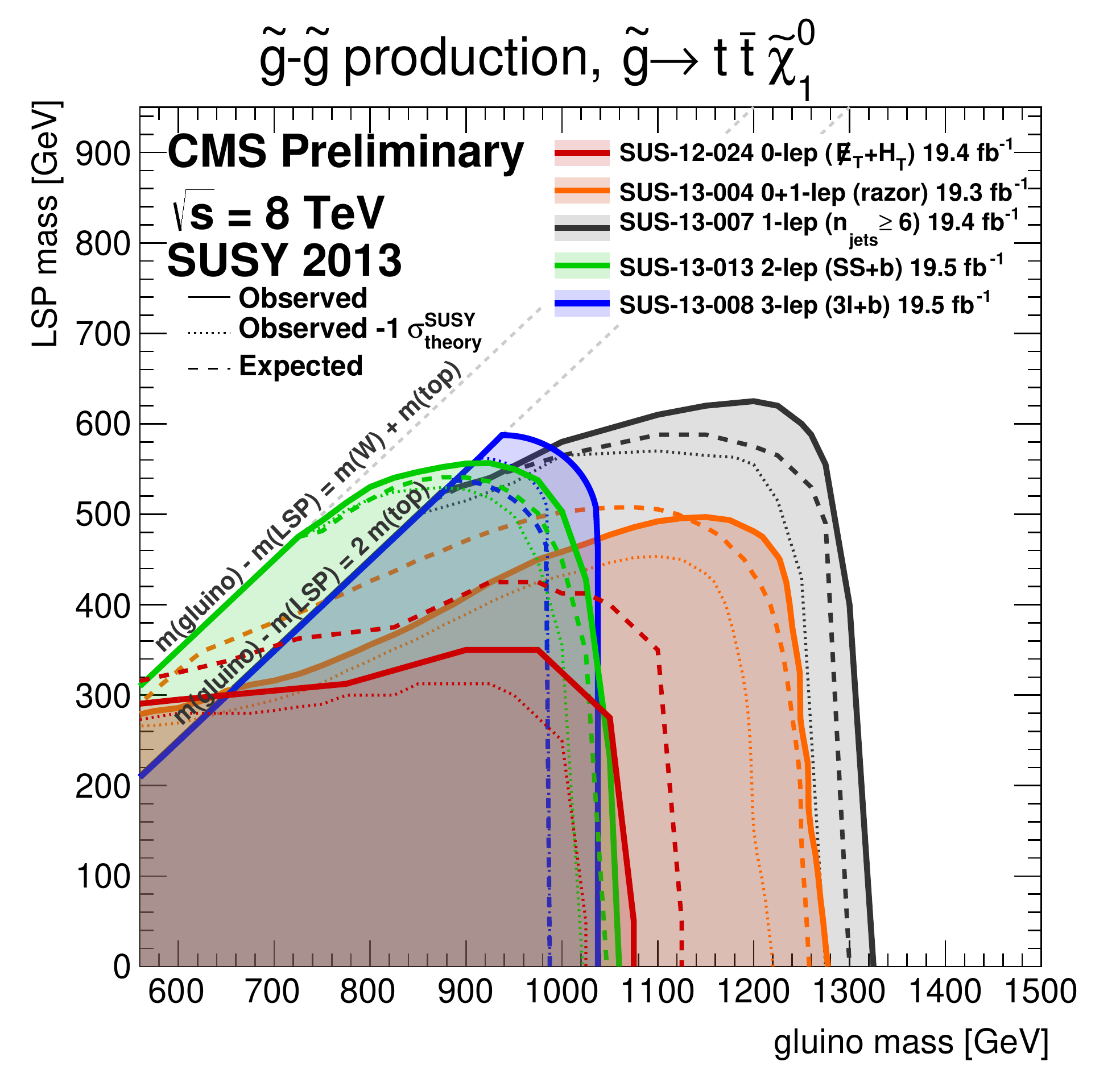}\hspace{2pc}%
\begin{minipage}[b]{0.4\textwidth}\caption{\label{fg:gl-ttlsp}Summary of observed and expected limits~\cite{cms-razor,cms-gluino,cms-ss2l} for gluino pair production with gluino decaying via a 3-body decay to a top, an anti-top and a neutralino. From Ref.~\cite{cms-susy-results}.}
\end{minipage}
\end{figure}

\begin{figure}[htb]
\includegraphics[width=0.5\textwidth]{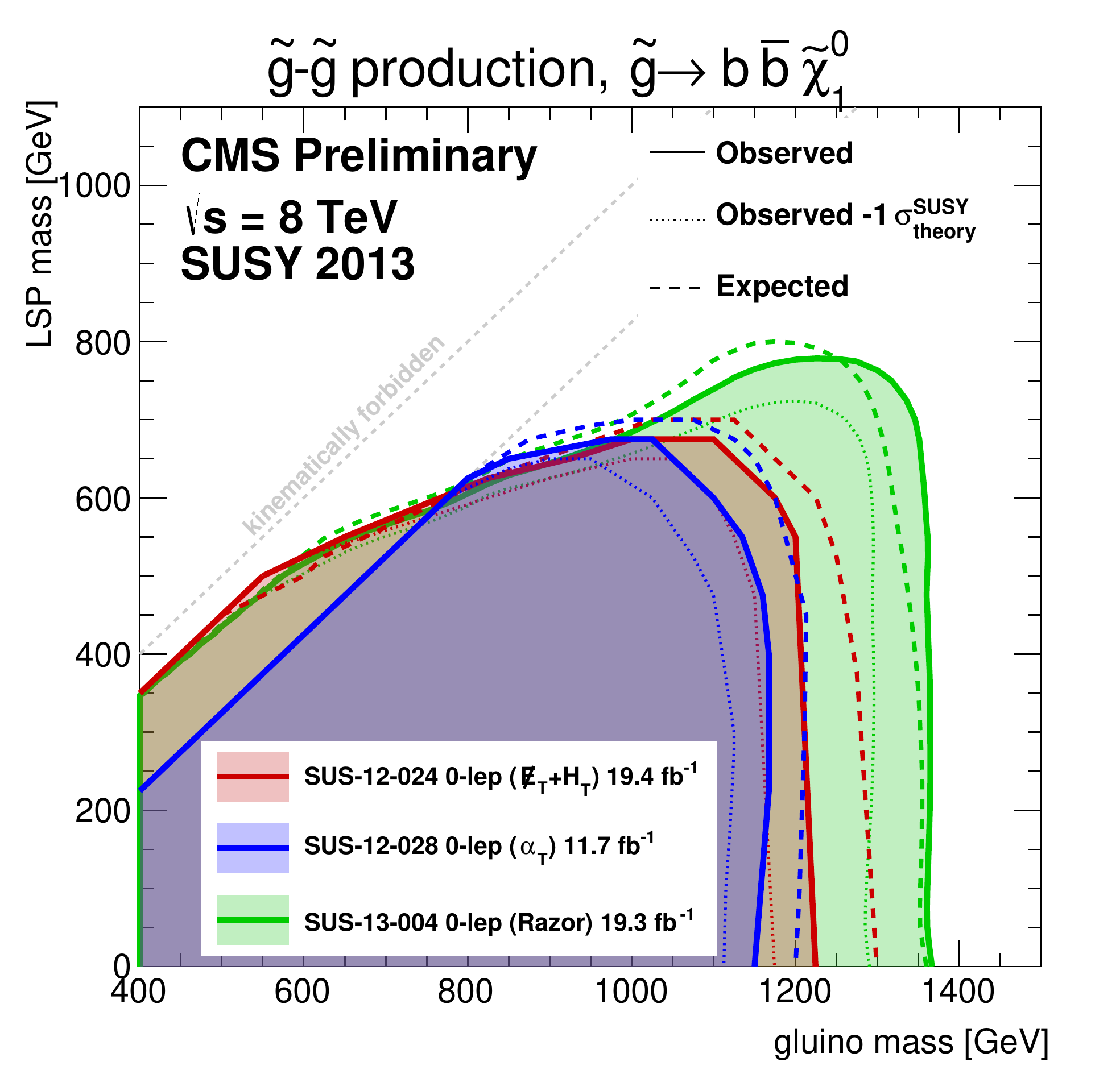}\hspace{2pc}%
\begin{minipage}[b]{0.4\textwidth}\caption{\label{fg:gl-bblsp}Summary of observed and expected limits~\cite{cms-razor,cms-gluino,cms-ss2l} for gluino pair production with gluino decaying via a 3-body decay to a bottom, an anti-bopttom and a neutralino. From Ref.~\cite{cms-susy-results}.}
\end{minipage}
\end{figure}

If the gluino is too heavy to be produced at the LHC, the only remaining possibility is the direct $\t{t}_1\t{t}_1$ and $\t{b}_1\t{b}_1$ production. If stop pairs are considered, two decay channels can be distinguished depending on the mass of the stop: $\t{t}_{1}\to b\t{\chi}_1^{\pm}$ and $\t{t}_{1}\to t\t{\chi}_1^0$. CMS and ATLAS carried out a wide range of different analyses in each of these modes at both 7~\tev\ and 8~\tev\ centre-of-mass energy. In all these searches, the number of observed events has been found to be consistent with the SM expectation. Limits have been set by ATLAS on the mass of the scalar top for different assumptions on the mass hierarchy scalar top-chargino-lightest neutralino~\cite{atlas-tt}, as shown in the left panel of Fig.~\ref{fg:directstop}. A scalar top quark of mass of up to 480~\gev\ is excluded at 95\% CL for a massless neutralino and a 150~\gev\ chargino. For a 300~\gev\ scalar top quark and a 290~\gev\ chargino, models with a neutralino with mass lower than 175~\gev\ are excluded at 95\% CL. 

\begin{figure}[htb]
\centerline{\includegraphics[width=0.5\textwidth]{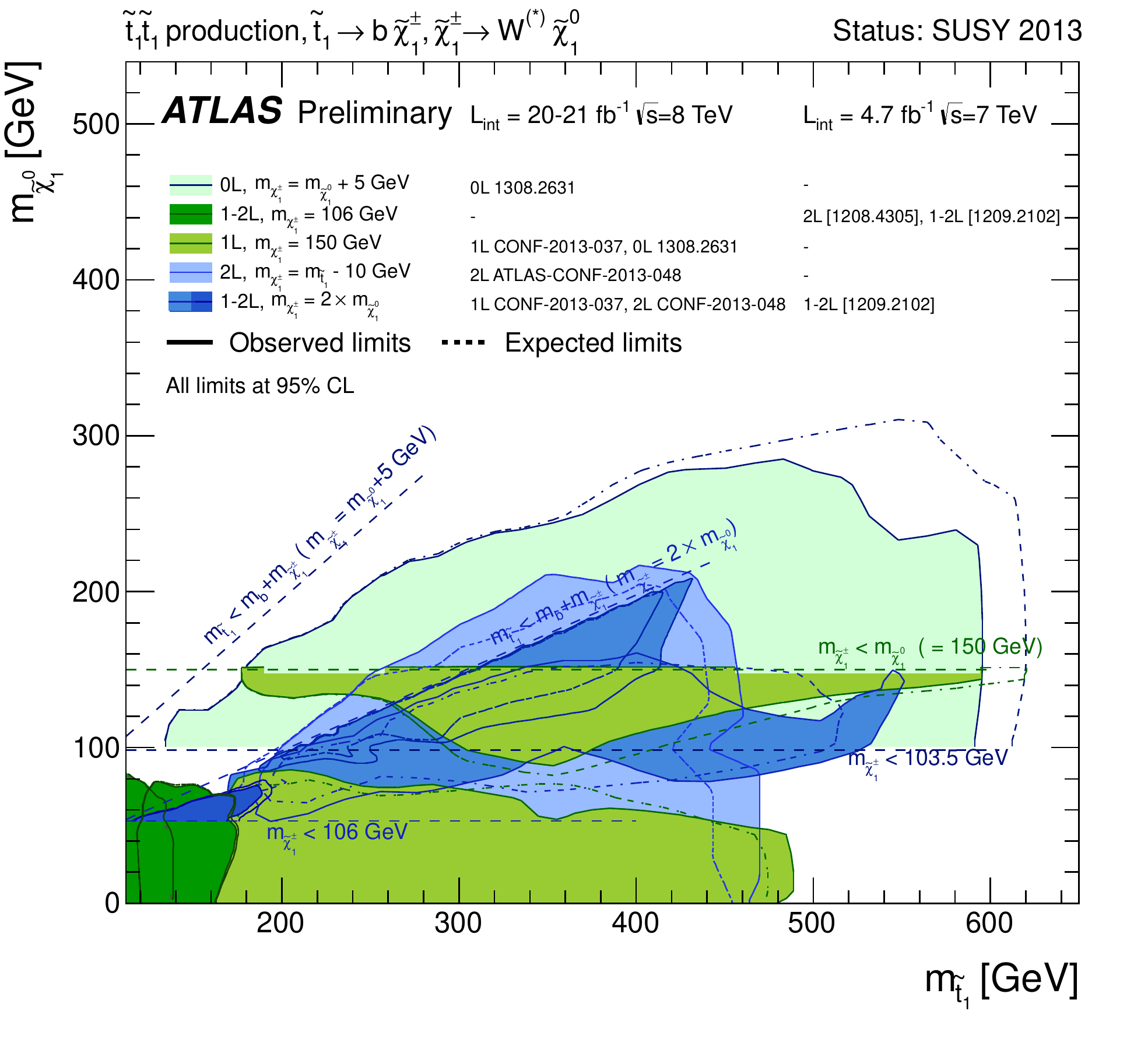}
  \hspace{0.0\textwidth}
  \includegraphics[width=0.5\textwidth]{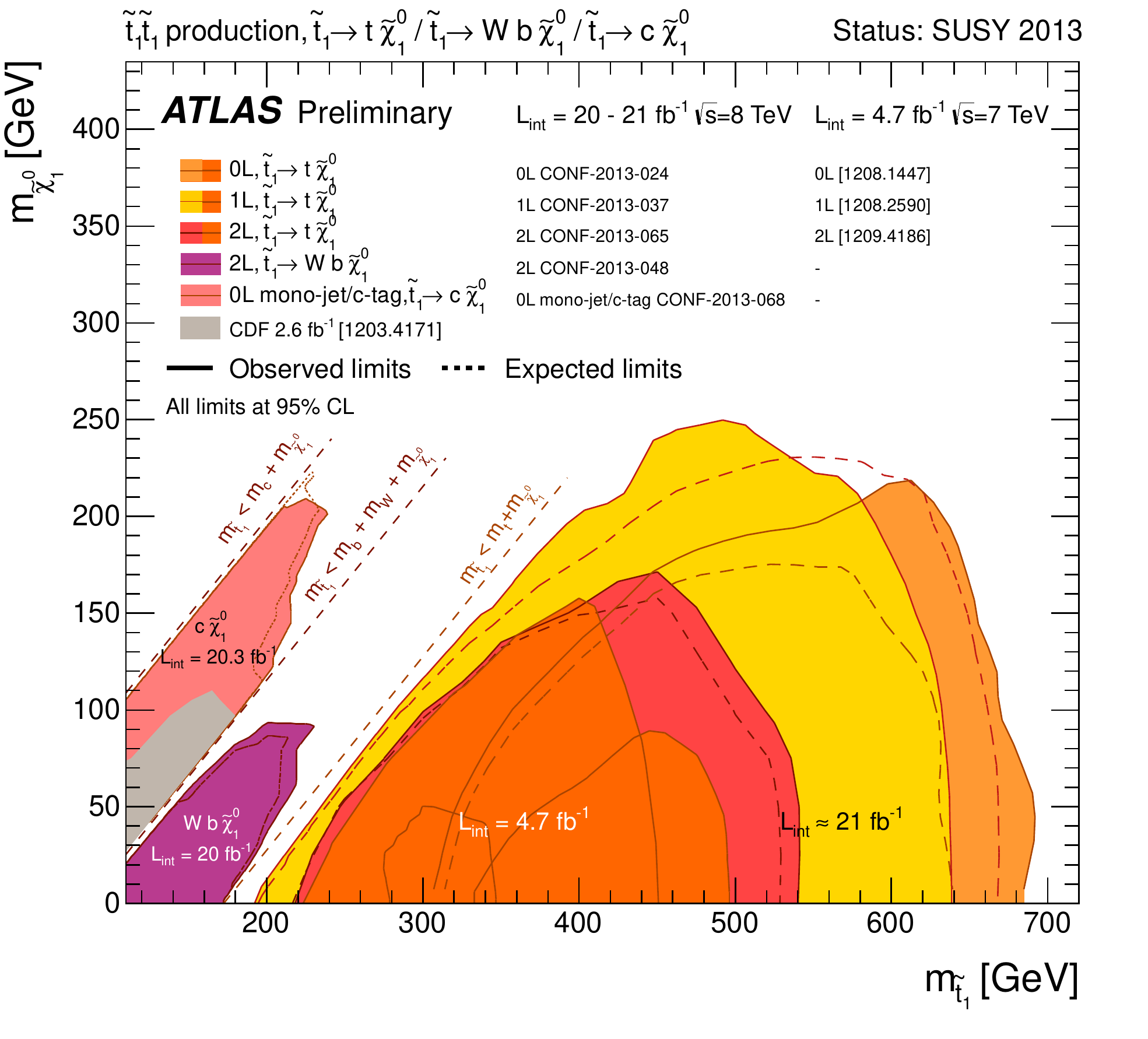}} 
\caption{Summary of the dedicated ATLAS searches~\cite{atlas-tt} for stop pair production based on $20-21~\ifb$ of $pp$ collision data taken at $\sqrt{s} = 8~\tev$, and $4.7~\ifb$ of $pp$ collision data taken at $\sqrt{s} = 7~\tev$. Exclusion limits at 95\% CL are shown in the $(\t{t}_{1},\,\t{\chi}_1^0)$ mass plane for channels targeting $\t{t}_{1}\to b\t{\chi}_1^{\pm},\:\t{\chi}_1^{\pm}\to W^{\pm}\t{\chi}_1^0$ decays (left) and $\t{t}_{1}$ decays to $t\t{\chi}_1^0$ or $W b \t{\chi}_1^0$ or $c \t{\chi}_1^0$ (right). The dashed and solid lines show the expected and observed limits, respectively, including all uncertainties except the theoretical signal cross section uncertainty. From Ref.~\cite{atlas-susy-results}.\label{fg:directstop}}
\end{figure}

For the case of a high-mass stop decaying to a top and neutralino ($\t{t}_{1}\to t\t{\chi}_1^0$), analyses requiring one, two or three isolated leptons, jets and large \met\ have been carried out. No significant excess of events above the rate predicted by the SM is observed and 95\% CL upper limits are set on the stop mass in the stop-neutralino mass plane. The region of excluded stop and neutralino masses is shown on the right panel of Fig.~\ref{fg:directstop} for the ATLAS analyses. Stop masses are excluded between 200~\gev\ and 680~\gev\ for massless neutralinos, and stop masses around 500~\gev\ are excluded along a line which approximately corresponds to neutralino masses up to 250~\gev. It is worth noting that a monojet analysis with $c$-tagging is deployed to cover part of the low-$m_{\t{t}_1}$, low-$m_{\t{\chi}_1^0}$ region through the $\t{t}_1\to c \t{\chi}_1^0$ channel.  

\subsection{Electroweak gaugino production}\label{sc:gaugino}

If all squarks and gluinos are above the TeV scale, weak gauginos with masses of few hundred gigaelectronvolts may be the only sparticles accessible at the LHC. As an example, at $\sqrt{s}Ê= 7~\tev$, the cross-section of the associated production $\t{\chi}_1^{\pm}\t{\chi}_2^0$ with degenerate masses of 200~\gev\ is above the 1-\tev\ gluino-gluino production cross section by one order of magnitude. Chargino pair production is searched for in events with two opposite-sign leptons and \met\ using a jet veto, through the decay $\t{\chi}_1^{\pm} \to \ell^{\pm}\nu\t{\chi}_1^0$. A summary of related analyses~\cite{cms-ewkino} performed by CMS is shown in Fig.~\ref{fg:ewkino}. Charginos with masses between 140 and 560~\gev\ are excluded for a massless LSP in the chargino-pair production with an intermediate slepton/sneutrino between the $\t{\chi}_1^{\pm}$ and the $\t{\chi}_1^0$. If $\t{\chi}_1^{\pm}\t{\chi}_2^0$ production is assumed instead, the limits range from 11 to 760~\gev. The corresponding limits involving intermediate $W$, $Z$ and/or $H$ are significantly weaker.  

\begin{figure}[htb]
\includegraphics[width=0.6\textwidth]{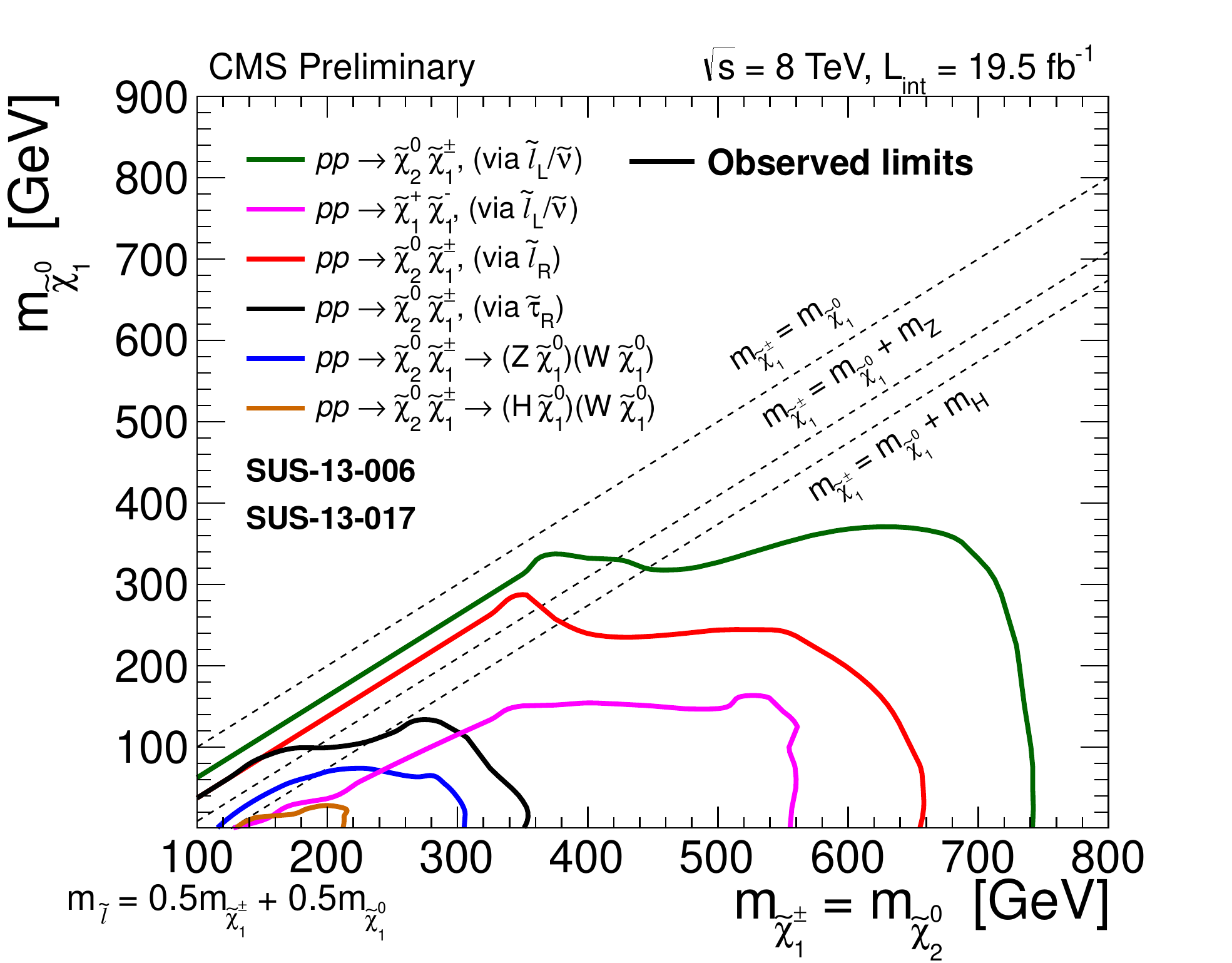}\hspace{1pc}%
\begin{minipage}[b]{0.3\textwidth}\caption{\label{fg:ewkino}Summary of observed limits for electroweak-gaugino production from CMS~\cite{cms-ewkino}. From Ref.~\cite{cms-susy-results}.}
\end{minipage}
\end{figure}

\subsection{$R$-parity violating SUSY and meta-stable sparticles}\label{sc:rpv}

\R-parity is defined as: $R = (-1)^{3(B-L)+2S}$, where $B$, $L$ and $S$ are the baryon number, lepton number and spin, respectively. Hence $R=+1$ for all Standard Model particles and $R=-1$ for all SUSY particles. It is stressed that the conservation of \R-parity is an \emph{ad-hoc} assumption. The only firm restriction comes from the proton lifetime: non-conservation of both $B$ and $L$ leads to rapid proton decay. \R-parity conservation has serious consequences in SUSY phenomenology in colliders: the SUSY particles are produced in pairs and the lightest SUSY particle is absolutely stable, thus providing a WIMP candidate. Here we highlight the status of RPV supersymmetry~\cite{rpv} searches at the LHC.

Both ATLAS and CMS experiments have probed RPV SUSY through various channels, either by exclusively searching for specific decay chains, or by inclusively searching for multilepton events. ATLAS has looked for resonant production of $e\mu$, $e\tau$ and $\mu\tau$~\cite{atlas-rpv-emu}, for multijets~\cite{atlas-rpv-multijets}, for events with at least four leptons~\cite{atlas-rpv-4l} and for excesses in the $e\mu$ continuum~\cite{atlas-rpv-emu-cont}. Null inclusive searches in the one-lepton channel~\cite{atlas-brpv} have also been interpreted in the context of a model where RPV is induced through bilinear terms~\cite{brpv}.

Recent CMS analyses are focused on studying the lepton number violating terms $\lambda_{ijk}L_iL_j\bar{e}_{k}$ and $\lambda'_{ijk}L_iQ_j\bar{d}_{k}$, which result in specific signatures involving leptons in events produced in $pp$ collisions at LHC. A search for resonant production and the following decay of $\t{\mu}$ which is caused by $\lambda'_{211}\neq0$ has been conducted~\cite{cms-rpv-ssmu}. Multilepton signatures caused by LSP decays due to various $\lambda$ and $\lambda'$ terms in stop production have been probed~\cite{cms-rpv-stop}. Reference~\cite{cms-rpv-4l} discusses the possibility of the generic model independent search for RPV SUSY in 4-lepton events. A summary of the limits set by several CMS analyses~\cite{cms-ss2l,cms-rpv-stop,cms-rpv} are listed in Fig.~\ref{fg:rpv}.

\begin{figure}[htb]
\centerline{\includegraphics[width=0.95\textwidth]{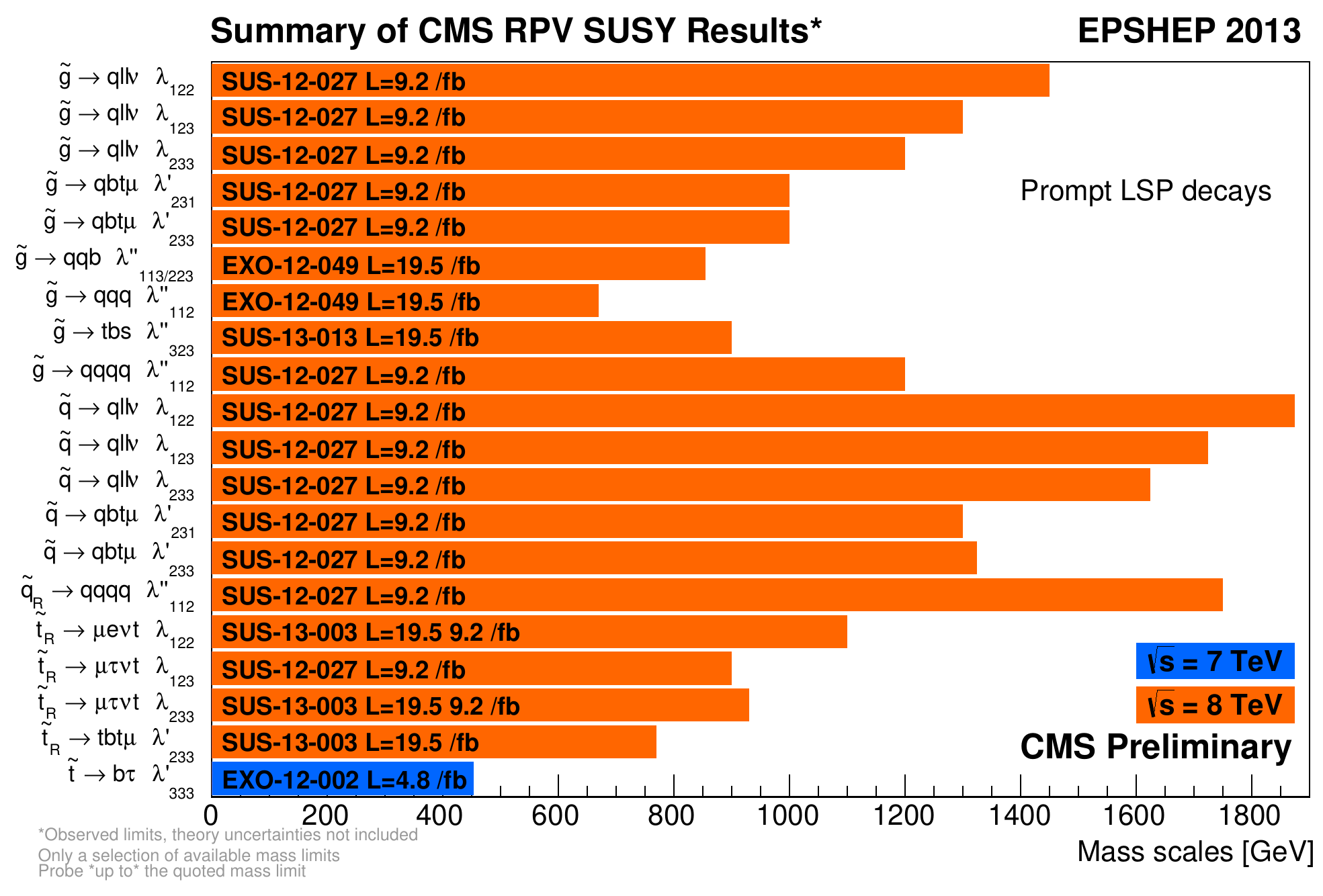}}
\caption{Best exclusion limits for the masses of the mother particles, for RPV scenarios, for each topology, for all CMS results~\cite{cms-ss2l,cms-rpv-stop,cms-rpv}. In this plot, the lowest mass range is $m_{\text{mother}}=0$, but results are available starting from a certain mass depending on the analyses and topologies. Branching ratios of 100\% are assumed, values shown in plot are to be interpreted as upper bounds on the mass limits. From Ref.~\cite{cms-susy-results}.\label{fg:rpv}}
\end{figure}

In view of the null results in other SUSY searches, it became mandatory to fully explore the SUSY scenario predicting meta-stable or long-lived particles. These particles, not present in the Standard Model, would provide striking signatures in the detector and rely heavily on a detailed understanding of its performance. In SUSY, non-prompt particle decay can be caused by (i) very weak RPV~\cite{atlas-dv}, (ii) low mass difference between a SUSY particle and the LSP~\cite{atlas-kinked}, or (iii) very weak coupling to the gravitino in GMSB models~\cite{rhadrons,nonp-phot}. A small part of these possibilities have been explored by the ATLAS~\cite{atlas-susy-results} and CMS~\cite{cms-susy-results} experiments covering specific cases, difficult to summarise here. There is still a wide panorama of signatures to be explored, in view of various proposed SUSY scenarios pointing towards this direction. 

As a last remark, we address the issue of (not necessarily cold) dark matter in RPV SUSY models. These seemingly incompatible concepts \emph{can} be reconciled in models with a gravitino~\cite{rpv-grav,trpv-gravitino} or an axino~\cite{rpv-axino,brpv-axino} LSP with a lifetime exceeding the age of the Universe. In both cases, RPV is induced by bilinear terms in the superpotential that can also explain current data on neutrino masses and mixings without invoking any GUT-scale physics~\cite{brpv}. Decays of the next-to-lightest superparticle occur rapidly via RPV interaction, and thus they do not upset the Big-Bang nucleosynthesis, unlike the \R-parity conserving case. Such gravitino DM is proposed in the context of $\mu\nu$SSM~\cite{munussm} with profound prospects for detecting $\gamma$ rays from their decay~\cite{munussm-dm}. 

Recent evidence on the four-year Fermi data that have found excess of a 130~\gev\ gamma-ray line from the Galactic Center~\cite{fermi130} have been studied in the framework of $R$-parity breaking SUSY. A decaying axino DM scenario based on the SUSY KSVZ axion model with the bilinear $R$-parity violation explains the Fermi 130~\gev\ gamma-ray line excess from the GC while satisfying other cosmological constraints~\cite{brpv-axino}. On the other hand, gravitino dark matter with trilinear RPV ---in particular models with the $LLE$ RPV coupling--- can account for the gamma-ray line, since there is no overproduction of anti-proton flux, while being consistent with big-bang nucleosynthesis and thermal leptogenesis~\cite{trpv-gravitino}.

\section{Summary and outlook}\label{sc:summary}

The nature of dark matter remains one of the mysteries of Particle Physics and Cosmology. Mono-$X$ searches at the LHC provide strong constraints on dark matter properties in an effective field theory formalism. Colliders are superior to direct searches if dark matter is very light ($< 10~\gev$) or if interactions are spin-dependent. Extensive efforts are currently in progress on the validity of the effective field-theory approach, the proper comparison with the results from direct detection experiments and the use of simplified models with light mediators. Analyses looking for specific models providing DM candidates, such as Supersymmetry, are ongoing. Searches continue with the full 2012 dataset but a new discovery might eventually require more energy and more data coming up in 2015.

\ack
The author is grateful to the XIV~MWPF organisers for the kind invitation and support that gave her the opportunity to present this plenary talk. She acknowledges support by the Spanish Ministry of Economy and Competitiveness (MINECO) under the projects FPA2009-13234-C04-01 and FPA2012-39055-C02-01, by the Generalitat Valenciana through the project PROMETEO~II/2013-017 and by the Spanish National Research Council (CSIC) under the JAE-Doc program co-funded by the European Social Fund (ESF). 


\end{document}